\documentstyle[12pt]{article}

\newcommand{\be}{\begin{equation}}
\newcommand{\ee}{\end{equation}}

\newcommand{\bbf}{\bf}
\newcommand{\ssl}{\sl}

\newcommand{\muo}{\mu_o}
\newcommand{\bea}{\begin{eqnarray}}
\newcommand{\eea}{\end{eqnarray}}

\begin{document}
\begin{center}
{\Huge The Semiclassical Instability of \\  
de Sitter Space}
  
\vspace{5 mm}  
  
{\bf Mikhail S. Volkov}  
  
\vspace{2 mm}  
  
{\it Laboratoire de Math\'ematiques  
et Physique Th\'eorique,\\  
Universit\'e de Tours,  
Parc de Grandmont,   
37200 Tours,   
FRANCE\\  
{\tt volkov@tpi.univ-tours.fr}  
}  
  
\end{center}  
  
\vspace{1 mm}  
\noindent  
{\sl  The effect of the spontaneous nucleation   
of black holes in de Sitter space is reviewed and the main  
steps of the calculation   
in {\em Nucl.Phys.}{\bbf B 582},~{\rm 313, 2000}   
of the one-loop amplitude of this process are summarized.   
The existence of such an effect   
suggests that de Sitter space is not   
a ground state of quantum gravity with a positive cosmological constant.%
\footnote{
Talk given at the workshop ``Quantum Gravity and Strings'',  
Dubna, June 2001.}   
}  
  
\vspace{10 mm}

The recent observational evidence for a positive cosmological constant   
has renewed interest in the dynamics of de Sitter space. 
In particular, quantum gravity in de Sitter space is attracting 
new attention (see for example \cite{Banks} and references therein). 
In this connection I would like to review 
in this lecture the effect of semiclassical instability of de Sitter
space with respect to spontaneous nucleation of black holes.   
Such an instability suggests that de Sitter space is not   
a ground state of quantum gravity with a positive cosmological constant
$\Lambda>0$.

The existence of the semiclassical   
instability in quantum gravity   
at finite temperature has been known for a long time.   
This is a non-perturbative effect mediated by certain gravitational instantons. 
In the $\Lambda=0$ case the effect was described in detail   
by  Gross, Perry and Yaffy \cite{Gross82}, while the case with   
$\Lambda>0$ was considered by  Ginsparg and Perry \cite{Ginsparg83}.  
The simplest way to understand this 
effect is to consider the semiclassical  partition function for 
Euclidean quantum gravity with $\Lambda>0$   
as the sum over gravitational instantons.    
It turns out that in this sum there is one 
term, determined by the $S^2\times S^2$ solution of the Euclidean   
Einstein equations,  
whose contribution is purely imaginary, because this   
(and only this) instanton  
 is not a local minimum
of the action. This renders the partition function and free energy 
complex, and the existence of an imaginary part of the free energy 
signals that the  system is metastable.    
Since the $S^2\times S^2$ instanton is the analytic continuation   
of the extreme Schwarzschild-de Sitter solution, one can  argue that 
this instability results in the nucleation of 
black holes in de Sitter space. The energy for this process comes from
the de Sitter heat-bath. 
  
This effect is inherently present in quantum gravity with $\Lambda>0$,  
since the value of the de Sitter temperature is fixed by $\Lambda$,  
and this gives rise to the thermal instability.   
For comparison, in the $\Lambda=0$ case the situation  
is different, since the temperature is then a free parameter   
and one can take the zero temperature limit. In this limit   
Minkowski space is stable \cite{Gross82}.

Summarizing, the  $S^2\times S^2$ instanton  
is analogues to the   
``bounce'' solution \cite{Callan77}, and so it is responsible for the   
formation of bubbles of a new phase.   
Its physical negative mode living in the physical TT-sector   
should not be confused with the   
negative modes from the conformal sector \cite{Gibbons78}.   
The latter exist for any background,  
and it seems that they do not imply any physical instabilities but emerge   
merely as a result of a bad  choice of variables in the path integral.   
It seems likely that   
if one was able to start from the path integral over the phase space  
and then make it covariant in some way, these modes would be absent  
\cite{Schleich87}.  Negative modes from the TT-sector are on the other hand  
physically significant. 
In fact, having factorized the conformal sector, it is only 
these remaining modes that render the partition function divergent.
This reflects the breakdown of the canonical ensemble for gravity    
due to black holes, whose degeneracy factor grows too fast with the energy.   
After complex-rotating   
the physical negative modes, the partition function becomes finite but complex,
thus again indicating the breakdown of the canonical description and the 
existence of metastability.    


A very nice feature of the $S^2\times S^2$ instanton is its high symmetry.   
This allows one to carry out  all the 
calculations exactly (in one loop), which is an   
exceptional situation in quantum gravity for non-trivial backgrounds.  
Such a calculation had not been done though until 
our recent paper with Andreas Wipf \cite{Volkov00}, where  
a complete derivation of the amplitude of the process was presented.   
In brief, we succeeded to explicitly determine the spectra of all the relevant  
fluctuation operators and to exactly compute their determinants within the   
$\zeta$-function scheme. 
Since calculations of this type are rather involved (and also rare), 
I would like
to briefly summarize below the essential steps of our procedure,
referring to \cite{Volkov00} for more details and references.

\section{Qualitative description} 
Let us consider the partition function for the gravitational
field
\be                   \label{r1}
Z=\int D[g_{\mu\nu}]\,{\rm e}^{-I}\, ,
\ee
where the integral is taken over Riemannian metrics with the
action
\be					\label{1}
I[g_{\mu\nu}]=-\frac{1}{16\pi G}\,\int_{\cal M} 
(R-2\Lambda) \sqrt{g}\, d^4x\, ,
\ee
with $\Lambda>0$. 
The extrema of this action are solutions $({\cal M},g_{\mu\nu})$ 
of the Euclidean Einstein equations
\be					\label{2}
R_{\mu\nu}=\Lambda g_{\mu\nu}\, .
\ee 
Such solutions are called instantons; all of them are compact. 
In the semiclassical approximation, for $G\Lambda\ll 1$,  
the path integral (\ref{r1}) 
is given by the sum over all instantons
\be            \label{r2}
Z\approx \sum_l Z_{[l]}\equiv
\sum_l\frac{\exp(-I_{[l]})}{\sqrt{{\rm Det}\Delta_{[l]}}}\,. 
\ee
Here $I_{[l]}$ is the action of the $l$-th instanton and  
the prefactor comes from the Gaussian integration over 
small fluctuations around this background, 
$\Delta_{[l]}$ being the corresponding fluctuation operator.

The leading contribution to this sum 
is given by those instantons whose action is minimal. 
One can expect that these will be the instantons whose symmetry is maximal. 
Among all solutions to equations (\ref{2}) there is one which 
is a maximal symmetry space -- this is the 
four-sphere with radius $\sqrt{3/\Lambda}$ and the standard metric
which can be written as 
\be                        \label{S4}
ds^2=(1-\frac{\Lambda}{3} r^2)\,dt^2
+\frac{dr^2}{1-\frac{\Lambda}{3}r^2}
+r^2(d\vartheta^2+\sin^2\vartheta d\varphi^2)\, ,
\ee
with $r\leq \sqrt{3/\Lambda}$. The action of this $S^4$ instanton 
is $I=-3\pi/\Lambda G$.  Hence, 
for $G\Lambda\ll 1$, the path integral (\ref{r1}) 
approximately is   
\be                \label{r3a}
Z\approx Z_{[S^4]}=
\frac{\exp(3\pi/\Lambda G)}{\sqrt{{\rm Det}\Delta_{[S^4]}}}. 
\ee
One can rewrite this expression as 
the partition function of some thermal system,
$Z={\rm e}^{-F/T}$.  Here the 
inverse temperature $1/T=2\pi\sqrt{\frac{3}{\Lambda}}$
is determined by the proper length of geodesics on $S^4$
(all of them are periodic),  
and up to subleading terms the free energy is
$F=-\frac{\sqrt{3}}{2G\sqrt{\Lambda}}$. 

One can wonder to which physical system refer these 
temperature and free energy ? In fact, 
they relate to the de Sitter space \cite{Gibbons77}. 
Upon analytic continuation $t\to it$ 
the metric 
(\ref{S4}) becomes the de Sitter metric 
restricted to the region inside the 
cosmological horizon, $r<r_{c}=\sqrt{3/\Lambda}$. 
Let us  call this region the Hubble region, or the causal diamond.   
De Sitter space has
the temperature $T$
determined by the surface gravity of the horizon, the entropy 
$S=\pi r_c^2/G$ 
determined by the area of the horizon, and the free energy $F=-TS$
contained inside the horizon \cite{Gibbons77,Gibbons77a}.  
The values of $T$ and $F$  
exactly coincide with those obtained above from the geometry of the 
$S^4$ instanton.   
  
Summarizing, the  
partition function of quantum gravity with 
$\Lambda>0$ and $G\Lambda\ll 1$ is determined by the 
thermodynamic parameters of de Sitter space, 
\be
Z\approx Z_{[S^4]}={\rm e}^{-F/T}\, . 
\ee

Let us now consider corrections to this formula due to other 
gravitational instantons contributing to (\ref{r2}):
\be
Z\approx {\rm e}^{-F/T}\left(1+\sum_{l\neq S^4}
\frac{Z_{[l]}}{Z_{[S^4]}}\right)\, . 
\ee
For $\Lambda G\ll 1$ all terms in this sum are exponentially
small and can be neglected, 
if only they are real. If there are complex terms, 
their contribution will be physically important despite their
smallness. Such complex terms can arise due to those 
instantons which are not local minima but saddle points of the action. 
If the number of their negative modes is odd,  
the determinants of the corresponding fluctuation operators will be 
negative, and the prefactors ${\sqrt{{\rm Det}\Delta_{[l]}}}$ in 
(\ref{r3a}) will therefore be complex.  

The question therefore arises: if among solutions of Eqs.(\ref{2})
there are those which are not local minima of the action ? 
A theorem quoted by Gibbons in Ref.\cite{Gibbons79LN} 
states that for $\Lambda>0$ there exists {\sl only one} such solution, 
which is the geometrical direct
product of $S^2\times S^2$ with the standard metric:
\be                                            \label{SS}
ds^2=\left.\left.\frac{1}{\Lambda}\right(
d{\vartheta_1}^2+
\sin^2{\vartheta_1}\,d{\varphi_1}^2+
d{\vartheta}^2+
\sin^2\vartheta\,d{\varphi}^2\right).
\ee
Its action is $I=-2\pi/G\Lambda$. 
As we will see, the solution has exactly one negative mode.
In view of this, the partition function can actually be approximated
by the semiclassical contributions of only two instantons, $S^4$
and $S^2\times S^2$: 
\be
Z\approx {\rm e}^{-F/T}\left(1+
\frac{Z_{[S^2\times S^2]}}{Z_{[S^4]}}\right)\approx
\exp\left\{-\frac{1}{T} \left 
(F-T\frac{Z_{[S^2\times S^2]}}{Z_{[S^4]}}\right)\right\}\, ,
\ee
where $Z_{[S^2\times S^2]}$ is purely imaginary.
Rewriting this
as $Z\approx {\rm e}^{-{\cal F}/T}$, the free energy ${\cal F}$ here 
will have the 
real part coinciding with the free energy $F$ 
in the Hubble region, and also the 
exponentially small imaginary part  
\be                      \label{ImF}
\Im ({\cal F})=-T\frac{Z[S^2\times S^2]}{ Z[S^4]}\, .
\ee
The existence of the imaginary part of the free energy indicates
that the system is metastable \cite{Langer67,Affleck81}.   
The conclusion therefore is that
de Sitter space, which is classically stable, becomes unstable 
semiclassically when the non-perturbative effects are taken
into account. 

The qualitative description of this instability 
was first given by Ginsparg and Perry \cite{Ginsparg83}, 
who argued that the quantum 
decay of de Sitter space will 
lead to the spontaneous nucleation of black holes. 
The logic is that, as the decay is the tunneling transition mediated by 
the $S^2\times S^2$ instanton, the structure of the configuration 
created during such a transition will be inherited from that  
of this instanton. The basic observation is then that the  $S^2\times S^2$
instanton can be obtained
via the analytic continuation of the Lorentzian 
Schwarzschild-de Sitter solution
\cite{Gibbons78b,Ginsparg83,Bousso96}
\be                           \label{SdS}
ds^2=-N\,dt^2
+\frac{dr^2}{N}
+r^2(d\vartheta^2+\sin^2\vartheta d\varphi^2)\, .
\ee
Here $N=1-\frac{2M}{r}-\frac{\Lambda}{3}r^2$, 
and for $9M^2\Lambda<1$ this function 
has roots at $r=r_{+}>0$ (black hole horizon) and
at $r=r_{++}>r_{+}$ (cosmological horizon). 
If one analytically continues via $t\to it$, the 
metric becomes Euclidean, 
but then one has to restrict the range of $r$ to $r_{+}<r<r_{++}$, 
since $N$
is negative otherwise. In addition, one has to 
identify $t$ with a suitable period, since otherwise the 
geometry will have conical singularities at  $r=r_{+}$ and 
$r=r_{++}$. In general one cannot remove both 
singularities at the same time, since the period of the   
identification of 
$t$ is determined by the surface gravity, which is different
for the two horizon. 
However, in the limit  
$r_{+}\to r_{++}\to\frac{1}{\sqrt{\Lambda}}$,
the surface gravities will be the same and both conical
singularities can be removed at the same time. Although
one might think that nothing will remain of the solution in this limit, 
this is not so. The limit $r_{+}\to r_{++}$ implies that 
$9M^2\Lambda=1-3\epsilon^2$ with $\epsilon\to 0$. 
One can introduce new coordinates  $\vartheta_1$ and 
$\varphi_1$ via
$\cos\vartheta_1=(\sqrt{\Lambda}r-1)/\epsilon+\epsilon/6$
and $\varphi_1=\sqrt{\Lambda}\,\epsilon\,\tau$. 
Passing to the new coordinates and taking the limit
$\epsilon\to 0$, the result is exactly the metric (\ref{SS}). 

The conclusion is that a tunneling transition via the
$S^2\times S^2$ instanton will create an extreme Schwarzschild-de Sitter
black hole. The radius of such a black hole is equal to the 
radius of the cosmological horizon, and so it will fill completely
the Hubble region.  However, 
the total volume of de Sitter space is infinite and it 
contains infinitely many Hubble regions. 
The black holes will emerge in some of these regions,  
but most of the regions will remain empty. The 
number of the filled regions divided by the number of the   
empty ones is the 
probability for the black hole nucleation in one region. This 
is proportional to $\Im({\cal F})$ in (\ref{ImF}).
As a result, black holes will nucleate with a certain probability all over
the space, like the bubbles in boiling water. 

The temperature of the nucleated black holes
can be 
read of from the $S^2\times S^2$ metric as the 
inverse length of the equator of any of the two spheres: 
$T_{\rm BH}=\frac{\sqrt{\Lambda}}{2\pi}$ (the same value can be
 obtained 
from the Lorentzian solution (\ref{SdS}) \cite{Bousso96}). 
This is {\sl different} from the temperature of the de Sitter heat bath.  
The origin of this discrepancy can be traced to 
the causal structure of de Sitter space: 
the fluctuations cannot {\sl absorb}  energy from 
{\sl and emit} energy into the 
whole of de Sitter space, but can only exchange  energy 
with  the Hubble region. 
Thus the energy exchange is restricted. As a result, the 
local temperature in the vicinity of a created defect may be
different from that of the heat bath. 

Using finally the classical formula of Langer and Affleck
\cite{Langer67,Affleck81}, the 
quasiclassical decay rate of de Sitter space  is
\be                      \label{rate}
\Gamma=2|\Im({\cal F})|=2T\,\frac{|Z_{[S^2\times S^2]}|}{Z_{[S^4]}}\, , 
\ee
where $T=\frac{1}{2\pi}\sqrt{\frac{\Lambda}{3}}$ is the temperature 
of the de Sitter heat bath. This formula 
gives the probability of a black hole nucleation per  
per Hubble volume and per unit time of a freely falling observer
\cite{Volkov00}. 

\section{Calculation of the path integral}
Let us now briefly describe the one-loop calculation of the 
partition functions $Z_{[S^2\times S^2]}$ 
and $Z_{[S^4]}$ entering formula (\ref{rate}).   
We use the standard Faddeev-Popov procedure and complex-rotate  
the conformal and TT-negative modes. Only the ghost operator has   
zero modes in its spectrum, which modes are    
associated with the background isometries. We integrate over these   
modes non-perturbatively, with the result (Eq.(\ref{killing}))  
being proportional to the volume   
of the isometry group.   
The main steps of the procedure are as follows.

For small fluctuations $h_{\mu\nu}$ 
around an instanton $({\cal M}, g_{\mu\nu})$
the action expands as $I[g_{\mu\nu}+h_{\mu\nu}]=I[g_{\mu\nu}]+
\delta^2 I+\ldots$~. Since $\delta^2 I$ has many zero modes
associated with diffeomorphisms, one has to add a gauge fixing term, 
which is chosen to correspond to the covariant background gauge  
\cite{Gibbons78a}:
\be					\label{13}
\delta^2 I_g=\gamma\left\langle
\nabla_\sigma h^\sigma_\rho-\frac{\gamma+1}{4\gamma}\,
\nabla_\rho h,
\nabla^\alpha h_\alpha^\rho-\frac{\gamma+1}{4\gamma}\,
\nabla^\rho h\right\rangle .
\ee
Here $\gamma$ is a real parameter and the scalar product 
is defined as 
$$
\langle h_{\mu\nu},h^{\mu\nu}\rangle=
\frac{1}{32\pi G}\,
\int_{\cal M} 
h_{\mu\nu}h^{\mu\nu}\sqrt{g}\, d^4 x;
$$
similarly for vectors and scalars. 
The one-loop partition function  then reads
\be					\label{17}
Z={\rm e}^{-I}\int D[h_{\mu\nu}]\,{\cal D}_{\rm FP}
\exp\left(-\delta^2 I-\delta^2 I_{g}\right), 
\ee
where the Faddeev-Popov factor is given by the integral over all
diffeomorphisms
\be					\label{18}
({\cal D}_{\rm FP})^{-1}=
\int D[\xi_\mu]\,
\exp\left(-\delta^2 I_{g}\right).
\ee

It is convenient to use the Hodge decomposition for fluctuations, 
\be					\label{4}
h_{\mu\nu}=\phi_{\mu\nu}+\frac14\,h\, g_{\mu\nu}+
\nabla_\mu\xi_\nu+\nabla_\nu\xi_\mu-
\frac12\,g_{\mu\nu}\nabla_\sigma\xi^\sigma\,,
\ee
where $\phi_{\mu\nu}$ is the transverse tracefree (TT) part, 
$\nabla_{\mu}\phi^\mu_\nu=\phi^\mu_\mu=0$, and 
$h$ is the trace. The 
longitudinal vector part also decomposes as 
\be                      \label{4a}  
\xi_\mu=\eta_\mu+\nabla_\mu\chi+\xi_{\mu}^{\rm H},  
\ee
where $\nabla_{\mu}\eta^\mu=0$ and the harmonic piece 
$\xi_{\mu}^{\rm H}$ vanishes for simply-connected
manifolds. With these decompositions the gauge fixing term 
and the gauge-fixed action 
$\delta^2 I_{gf}=\delta^2 I+\delta^2 I_{g}$ are diagonal:
\bea					\label{15}
&&\delta^2 I_g=\gamma\langle\eta_\mu,\Delta_1^2\eta^\mu\rangle
+\frac{1}{16\gamma}\langle
(\tilde{h}+2\tilde{\Delta}^\gamma_0\chi),
\Delta_0 (\tilde{h}+2\tilde{\Delta}^\gamma_0\chi)\rangle\, . \\
&&\delta^2 I_{gf}=\frac12\,              \nonumber 
\langle\phi^{\mu\nu},\Delta_2\phi_{\mu\nu}\rangle
+\gamma\langle\eta_\mu,\Delta_1^2\eta^\mu\rangle 
+\frac14\,\langle\chi,\Delta_0\tilde{\Delta}_0
\tilde{\Delta}_0^\gamma\chi\rangle
-\frac{1}{16\gamma}\,\langle h,\tilde{\Delta}_0
h\rangle\, . \nonumber 
\eea
Here the tensor fluctuation operator is
\be					\label{8}
\Delta_2\phi_{\mu\nu}=-\nabla_\sigma\nabla^\sigma
\phi_{\mu\nu}
-2R_{\mu\alpha\nu\beta}\phi^{\alpha\beta}\, ,
\ee
the vector operator and scalar operators are
\be					\label{9}
\Delta_1=-\nabla_\sigma\nabla^\sigma-
\Lambda\, ,~~~~~\Delta_0=-\nabla_\sigma\nabla^\sigma\, ,
\ee
while $\tilde{\Delta}_0=3\Delta_0-4\Lambda$ and 
$\tilde{\Delta}^{\gamma}_0=\gamma\tilde{\Delta}_0-\Delta_0$  
also $\tilde{h}=h-2\nabla_\mu\xi^\mu$. 

To compute the path integrals in  (\ref{17}),(\ref{18}), 
all fields are expanded with respect to the bases associated with complete sets 
of eigenstates of the operators $\Delta_2$, $\Delta_1$ and $\Delta_0$
with some Fourier coefficients $C$. For example,  
$h=\sum_n C_n h_n$ where
$\Delta_0\,h_n=\lambda_n\,h_n$.  
The quadratic actions (\ref{15}) then reduces to quadratic forms in the 
coefficients $C$, 
and the path-integration in (\ref{17}),(\ref{18}) is performed 
by integrating over all $dC$. 
The perturbative path integration measure is defined as 
the square root of the determinant of the 
metric on the function space
of fluctuations:
\be                                    \label{32}
D[h_{\mu\nu}]\sim\sqrt{{\rm Det}(\langle dh_{\mu\nu}, 
dh^{\mu\nu}\rangle})\, ,\ \ \ \ \
D[\xi_{\mu}]\sim\sqrt{{\rm Det}(\langle d\xi_{\mu}, 
d\xi^{\mu}\rangle})\, ,\ \ 
\ee
where the differentials refer to the Fourier 
coefficients $C$ of the expansions of $h_{\mu\nu}$ and
$\xi_{\mu}$.  For example, $dh=\sum_n dC_n h_n$. 
The normalization is chosen such that
$\int D[h_{\mu\nu}]\exp\left(-\frac{\mu_o^2}{2}\, 
\langle h_{\mu\nu},h^{\mu\nu}\rangle\right)=1$
and 
$
\int D[\xi_\mu]\exp\left(-\mu_o^4\, 
\langle \xi_{\mu},\xi^{\mu}\rangle\right)=1
$,
where $\mu_o$ is an arbitrary renormalization 
parameter with the dimension of an inverse length.

Using (\ref{4}),(\ref{4a}) the metrics on the space of fluctuations are
\bea
\langle dh_{\mu\nu},dh^{\mu\nu}\rangle&=&
\langle d\phi_{\mu\nu},d\phi^{\mu\nu}\rangle
+2\langle d\eta_\mu,\Delta_1 d\eta^\mu\rangle
+\langle d\chi,\Delta_0\tilde{\Delta}_0 d\chi\rangle
+\frac14\,\langle dh,dh\rangle\, ,\nonumber \\
\langle d\xi_{\mu},d\xi^{\mu}\rangle&=&
\langle d\eta_\mu,d\eta^\mu\rangle
+\langle d\chi,\Delta_0 d\chi\rangle\, .   \label{33}
\eea

The path integrals in  (\ref{17}),(\ref{18}) reduce then to 
infinite products of ordinary integrals over the Fourier coefficients $C$. 
Most of these integrals are Gaussian,  and their computation gives 
products of eigenvalues $\sigma_s$ for coexact vectors 
and eigenvalues  $\epsilon_k$ for TT-tensors in the resulting formula 
for $Z$ in Eq.(\ref{46}) below. 
There are, however, also non-Gaussian integrals, which 
can be of three different types. 

$\bullet$ First, for the $S^2\times S^2$ instanton there exists a 
TT tensor mode with a negative eigenvalue $\epsilon_{-}<0$. 
The integration
over this mode is carried out with the complex contour rotation 
\cite{Callan77} leading to the complex factor 
$\Omega_{\rm neg}=\mu_o/(2i\sqrt{|\epsilon_{-}|})$ in (\ref{46}).
For any other instanton $\Omega_{\rm neg}=1$. 

$\bullet$ Next, for any instanton there are infinitely many conformal 
negative modes associated with the `wrong' sign term 
$-\frac{1}{16\gamma}\,\langle h,\tilde{\Delta}_0
h\rangle$ in (\ref{15}). This is the manifestation of the well-known 
problem of conformal sector in Euclidean quantum gravity 
\cite{Gibbons78}. Although its real understanding 
is lacking,  it seems that this problem is essentially an artifact of 
the bad choice of variables in the path integral \cite{Schleich87}. 
The prescription is then to perform the complex rotation 
$h\to ih$ in the space spanned by positive eigenstates of the operator 
$\tilde{\Delta}_0$ \cite{Volkov00}.   After this the quadratic forms become
positive-definite and the integrals can be computed. Remarkably,
the resulting effect of conformal modes is then exactly cancelled by the 
contribution of the exact parts $\chi$ of the 
vectors.   There is only one scalar mode
which contributes to the final answer:  the constant 
conformal mode present for any background and giving the factor 
$\Omega_{\rm conf}=\frac{\mu_o}{\sqrt{2\Lambda}}$ to (\ref{46}).
Only for the $S^4$ instanton there are additional 5 conformal
Killing scalars which also contribute, and so the answer in the $S^4$ 
case is $\Omega_{\rm conf}=\frac{\mu_o}{\sqrt{2\Lambda}}
\left(\sqrt{\frac{2\Lambda}{3}}\frac{1}{\mu_o}\right)^5$.

$\bullet$ Finally, after fixing the gauge, there remain a finite number of 
zero modes of the vector operator $\Delta_1$. These are
associated with the {\sl background isometries}. Such modes 
are not contained in the fluctuation measure 
$D[h_{\mu\nu}]$, since they do not contribute to the metric
$\langle dh_{\mu\nu},dh^{\mu\nu}\rangle$ in (\ref{33}), but 
they {\sl are contained} in the ghost measure $D[\xi_{\mu}]$ and   
should therefore be taken into account. In fact, the effect of these   
modes turns out to be very important. 
In treating these modes, we follow the approach of  t'Hooft 
\cite{tHooft76} and Osborn \cite{Osborn81}, whose idea is to carry out
the integration over these modes 
non-perturbatively, which amounts to 
integrating over the isometry group ${\cal H}$. 
 The corresponding integration
measure must be proportional to the Haar measure. If 
${\cal H}$ acts on ${\cal M}$ via 
${x}^\mu\to {x}^\mu(C_j)$, where $j=1,\ldots {\rm dim}{\cal H}$,
the vector zero modes are the Killing vectors 
$K_j=\frac{\partial}{\partial C_j}\equiv
\frac{\partial x^\mu}{\partial C_j}\frac{\partial}{\partial x^\mu}$,
and the integration over these modes gives 
$
\Omega_{\rm Killing}=\int
\left(\prod_j
\frac{\mu_o^2}{\sqrt{\pi}}\,
\left|\left|\frac{\partial}{\partial C_j}
\right|\right|\right)
d\mu(C)\, .   
$  
Here $\frac{\partial}{\partial C_j}$ is computed at $C_j=0$
and the normalization of the Haar measure 
$d\mu(C)$ of ${\cal H}$ is fixed by the
condition that at the unity, $C_j=0$, 
the perturbative measure is reproduced:
$d\mu(C)\to \prod_j dC_j$ as $C_j\to 0$. 
For the $S^2\times S^2$ instanton the isometry group is 
${\cal H}=SO(3)\times SO(3)$, while in the $S^4$ case 
${\cal H}=SO(5)$. This gives, respectively \cite{Volkov00}
\be                                       \label{killing}
\Omega_{\rm Killing}=\frac{64\pi^4(\muo)^{12}}
{27(\Lambda G)^3},~~~
\Omega_{\rm Killing}=\left(\frac{9}{10}\right)^5\,\frac{128\pi^6}{3}\,
\frac{(\muo)^{20}}{(\Lambda G)^5}. 
\ee

All this finally leads to the following expression for the one-loop
partition function of small fluctuations around a given instanton
background:
\be					\label{46}
Z=
\frac{\Omega_{\rm neg}\Omega_{\rm conf}}{\Omega_{\rm Killing}}\,
\left(\prod_{\sigma_s>0}
\frac{\sqrt{\sigma_s}}{\mu_o} \right)\left(
\prod_{\epsilon_k>0}
\frac{\mu_o}{\sqrt{\epsilon_k}}\right){\rm e}^{-I},
\ee
where all dependence on the gauge parameter $\gamma$
has disappeared. 

In order to use this formula, we need to explicitly know
the spectra of the tensor fluctuation operator (\ref{8}) on the 
space of the TT-tensors subject to 
$\nabla_\mu\phi^\mu_\nu=\phi^\mu_\mu=0$, and also those for the 
vector operator (\ref{9}) on the space of coexact vectors 
$\nabla_\mu\eta^\mu=0$. For the 
$S^2\times S^2$ instanton one finds \cite{Volkov00}\\
\vspace{3 mm}  
\begin{tabular}{|c|c|c|c|}\hline
operator & eigenvalue & degeneracy &  \\ \hline
$\Delta_2$& $-2\Lambda$ & $1$ &  \\
& $2\Lambda$  & $9$ &   \\
& $(j(j+1)-2)\Lambda$ & $2(2j+1)$ & $j\geq 2$ \\
& $j(j+1)\Lambda$ & $18(2j+1)$ & $j\geq 2$ \\
& $(j_1(j_1+1)+j_2(j_2+1)-2)\Lambda$ &
$5(2j_1+1)(2j_2+1)$  &
$j_1,j_2\geq 2$  \\
\hline
$\Delta_1$ & 0 & 6 &  \\
& $(j(j+1)-2)\Lambda$ &
$2(2j+1)$  &
$j\geq 2$  \\
& $(j_1(j_1+1)+j_2(j_2+1)-2)\Lambda$ &
$3(2j_1+1)(2j_2+1)$  &
$j_1,j_2\geq 1$  \\
\hline
$\Delta_0$ & $(j_1(j_1+1)+j_2(j_2+1))\Lambda$ &
$(2j_1+1)(2j_2+1)$  &
$j_1,j_2\geq 0$  \\
\hline
\end{tabular}\\
These spectra contain one negative tensor mode
and six vector zero modes associated with the $SO(3)\times SO(3)$
isometry group; all other modes are positive. 
For the $S^4$ instanton one has 
\begin{center}
\begin{tabular}{|c|c|c|c|}\hline
operator & eigenvalue & degeneracy &  \\ \hline
$\Delta_2$& $\frac{\Lambda}{3}j(j+3)$ & 
$\frac56(j-1)(j+4)(2j+3)$ & $j\geq 2$ \\
\hline
$\Delta_1$& $\frac{\Lambda}{3}(j(j+3)-4)$ & 
$\frac12 j(j+3)(2j+3)$ & $j\geq 1$ \\
\hline
$\Delta_0$& $\frac{\Lambda}{3}j(j+3)$ & 
$\frac16(j+1)(j+2)(2j+3)$ & $j\geq 0$ \\
\hline
\end{tabular}
\end{center}  
In this case all eigenvalues are positive, apart from  
ten vector zero modes which are generators of SO(5). 

The next step is to compute the infinite products of these eigenvalues
in Eq.(\ref{46}). 
This can be done with the use of the $\zeta$-function regularization,
in which scheme the product of an infinite discrete 
set of numbers $\sigma_s$
by definition is
\be                    \label{a42}
\prod_s
{\frac{{\sigma_s}}{\mu}} 
=\exp\left\{-\zeta^\prime(0)
-\zeta(0)\ln\mu\right\}\, ,    
\ee
where the $\zeta$-function 
\be                   \label{a43}
\zeta(z)=\sum_s (\sigma_s)^{-s}\, .
\ee
The spectra in the tables above lead to the $\zeta$-functions
of the following three basic types
\bea                                   \label{A1}
Z(k,\nu|z)&=&\sum_{n=k}^\infty 
\sum_{m=k}^\infty\,\frac{(2n+1)\,(2m+1)}
{\left\{(2n+1)^2+(2m+1)^2+\nu\right\}^z}\, , \\
{\zeta}(k,\nu|z)&=&\sum_{n=k}^\infty 
\frac{(2n+1)}
{\left\{(2n+1)^2+\nu\right\}^z}\, ,\nonumber \\
{\cal Q}(k,\nu,c|z)&=&\sum_{j=k}^\infty
\frac{(2j+3)(j(j+3)+c)}{\{j(j+3)+\nu\}^z}\,,
\eea  
where $\Re(z)$ must be large enough to ensure convergence. 
The analytic continuation of these
expressions to arbitrary values of $z$ has been carried out
in \cite{Volkov00}. This gives for $z=0$
\bea
Z(k,\nu|0)&=&
\frac{1}{32}\,\nu^2-\frac{1}{24}\,\nu+\frac{1}{2}\, k^2\nu
+2k^4-\frac23\,k^2+\frac{13}{360}\, , \nonumber \\
{\zeta}(k,\nu|0)&=&\frac{1}{12}-\frac14\,\nu-k^2\, , \\
{\cal Q}
(k,\nu,c|0)&=&-\frac{k^4}{2}-2k^3-(2c+1)\frac{k^2}{2} 
+(3-2c)\,k+\frac{\nu^2}{2}+(4-3\nu)\frac{c}{3}
-\frac{11}{15}\, ,  \nonumber 
\eea
which quantities define the `regularized numbers of eigenvalues' of the 
operators and determine the anomalous scaling behavior of the 
determinants. The values of the determinants themselves are given  
by the derivatives of the $\zeta$-functions at zero, which 
can be expressed in quadratures and computed numerically. 
The values
needed in Eq.(\ref{46}) are $Q'(2,0,-4|0)=3.72344$, 
$Q'(2,-4,0|0)=6.65246$, and  $Z'(2,-10|0)=-18.3118$ \cite{Volkov00}. 

Collecting everything together, 
the one-loop partition functions in the $\zeta$-function   
regularization scheme are 
\be                     \label{a57}
Z_{[S^2\times S^2]}=
-i\, 0.3667\times (\Lambda G)^3
\muo^{-\frac{98}{45}}
\exp\left(\frac{2\pi}{\Lambda G}\right)
\ee
and
\be                 
Z_{[S^4]}=
0.0047\times (\Lambda G)^5
\muo^{-\frac{571}{45}}
\exp\left(\frac{3\pi}{\Lambda G}\right).  \label{a57a}
\ee  
Up to my knowledge, such closed expressions were obtained   
for the first time in \cite{Volkov00}.   
Here the numerical prefactors are determined by the $\zeta$-regularized  
determinants, the factors of $\Lambda G$ come from the background isometries  
(see Eq.(\ref{killing})).   
The powers of $\muo$ are the anomalous dimensions, they receive   
contributions from all modes, including the isometries.   
As is seen from Eq.(\ref{killing}), the anomalous effect of the isometries  
is considerable, and had it been neglected the result would be very different.     
The anomalous dimension $-\frac{571}{45}$  
for $S^4$ in (\ref{a57a}) agrees with the analysis of   
\cite{Christensen80,Fradkin84}.

The last step is to 
insert into Eq.(\ref{rate}) to find 
\bea                     
\Gamma=
14.338\,\sqrt{\Lambda}\, 
(G\Lambda)^{-2} (\mu_o\Lambda)^{\frac{473}{45}}
\exp\left(-\frac{\pi}{\Lambda G}\right).         \label{final}
\eea
This gives the probability of the spontaneous nucleation of a black hole  
inside the finite volume enclosed by the de Sitter cosmological horizon   
per unit time of a freely falling observer. The formula applies for   
$\Lambda G\ll 1$. Due to non-renormalizability of gravity, the cutoff  
parameter $\muo$ remains undetermined; for estimates one can for example  
set $\muo=G$. The subsequent real time evolution of the created black holes  
is an issue requiring special study \cite{Bousso99}.

\vspace{5 mm}  
  
\noindent  
  

\end{document}